\documentclass[pre,twocolumn,twoside,showpacs,superscriptaddress,floatfix,eqsecnum]{revtex4}
\usepackage{graphics,amssymb,amsmath,epsfig,color}
\epsfclipon

\begin{document}
\title{Finite-size scaling, dynamic fluctuations, and hyperscaling relation in the Kuramoto model}
\author{Hyunsuk Hong}
\affiliation{Department of Physics and Research Institute of Physics and Chemistry, Chonbuk National University, Jeonju 561-756, Korea}
\author{Hugues Chat\'e}
\affiliation{Service de Physique de l'Etat Condens\'e, CEA-Saclay, CNRS UMR 3680,
91191 Gif-sur-Yvette, France}
\affiliation{Beijing Computational Science Research Center, Beijing 100084, China}
\author{Lei-Han Tang}
\affiliation{Department of Physics, Hong Kong Baptist University, Kowloon Tong,
Hong Kong SAR, China}
\affiliation{Beijing Computational Science Research Center, Beijing 100084, China}
\author{Hyunggyu Park}
\affiliation{School of Physics and QUC, Korea Institute for Advanced Study, Seoul
130-722, Korea}
\date{\today}

\begin{abstract}
We revisit the Kuramoto model to explore the finite-size scaling (FSS) of the order parameter
and its dynamic fluctuations
near the onset of the synchronization transition, paying particular attention to
effects induced by the randomness of the intrinsic frequencies of oscillators. For a population of size $N$,
we study two ways of sampling the intrinsic frequencies according to the {\it same}
given unimodal distribution $g(\omega)$.
In the `{\em random}' case, frequencies are generated independently in accordance with
$g(\omega)$, which gives rise
to oscillator number fluctuation within any given frequency interval.
In the `{\em regular}' case, the $N$ frequencies are generated in a deterministic
manner that minimizes the oscillator number fluctuations, leading to
quasi-uniformly spaced frequencies in the population.  We find that the
two samplings yield substantially different finite-size properties with
clearly distinct scaling exponents. Moreover, the hyperscaling relation
between the order parameter and its fluctuations is valid in the regular case,
but is violated in the random case. In this last case, a self-consistent mean-field theory that completely ignores
dynamic fluctuations correctly predicts the FSS exponent of the order parameter but not its critical amplitude.

\date{\today}
\end{abstract}
\pacs{89.75.-k, 05.45.Xt, 05.45.-a}
\maketitle

\section{Introduction}

We revisit the Kuramoto model~\cite{ref:Kuramoto} to
investigate the phase order parameter and its dynamic fluctuations
near the onset of the synchronization transition. In particular, we focus
on  their critical finite-size scaling (FSS) behavior.
Even though the Kuramoto model and its extended versions have been widely studied~\cite{ref:synchronization,ref:Acebron},
some issues persist, in particular about the above points.
In the few existing studies~\cite{ref:Daido,ref:Hildebrand},
the critical properties near the transition point have only been partially resolved, which prompted the present work.

FSS is particularly important for data analysis, since real systems are always finite, and particularly so in many experimental systems,
biological or not~\cite{ref:Kiss}.
Recently, the FSS of the synchronization order parameter in the Kuramoto model and several extended versions
on complex networks has been investigated. An unusual FSS exponent
$\bar\nu=\frac{5}{2}$ has been found~\cite{ref:HPC_PRE,ref:HCPT_PRL,ref:HPT_Synch_SFN}.
This value was obtained in the classic case,  hereafter referred to as the {\it random distribution},
where the individual frequencies of oscillators are sampled
independently according to a given unimodal distribution $g(\omega)$.
In such a situation, oscillator frequencies can be arbitrarily close to each other even in a finite population.
Interestingly, when the oscillator density fluctuation along the frequency axis is suppressed in
a population with {\it regular distribution}, a very different FSS exponent $\bar\nu=\frac{5}{4}$ is obtained.
Here we present details of our numerical study that led to this previously reported value~\cite{ref:Tang,ref:HUP}.

Dynamic fluctuations of the order parameter in the Kuramoto model is another
long-standing problem~\cite{ref:Strogatz,ref:Acebron}.
In the perturbation theory developed by Daido~\cite{ref:Daido}, it is claimed
that they scale differently when approaching the synchronization threshold from above and from below,
with the scaling exponent $\gamma=\frac{1}{4}$ in the supercritical region and $\gamma^\prime=1$ in the
subcritical region.
This result is quite surprising by itself, in the sense that
conventional scaling theory predicts identical scaling behavior on either side of the transition.
Note that Daido used the regular frequency distribution in numerical investigations as well
as in his perturbation theory~\cite{ref:Daido}.

In this paper, we report extensive numerical results on the FSS of the order parameter
and its dynamic fluctuation in the random and regular realizations of the intrinsic frequency
distribution. The FSS exponent in the regular case is estimated numerically as
$\bar\nu\simeq \frac{5}{4}$, half the value of its random counterpart.
In contrast to Daido's predictions, we find that dynamic fluctuations
scale with the same exponent $\gamma=\gamma'$ on both sides of the transition.
Interestingly, our data supports $\gamma\simeq\gamma^\prime\simeq 1$ for the random frequency distribution
and $\gamma\simeq\gamma^\prime\simeq \frac{1}{4}$ for the regular case.
Given the order parameter exponent $\beta=\frac{1}{2}$,
the hyperscaling relation $\gamma=\bar\nu-2\beta$ is thus violated in the random case, but obeyed
in the regular case.
We finally briefly discuss the implications of our observations on the ordering process and dynamic response in the Kuramoto model.

The paper is organized as follows.  In Sec.~II, we introduce the Kuramoto model and present the two types of frequency realizations that we consider in this paper.
In Sec.~III, we investigate the FSS of the order parameter and its dynamic fluctuations with the random frequency distribution
and discuss the validity of the hyperscaling relation in the model.
In Sec.~IV, we repeat the investigation and analysis with the regular distribution.
We summarize and discuss the results in Sec.~V. Appendix A presents the mean-field approximation to
the FSS in the case of the random frequency distribution.

\section{The Kuramoto model}

The Kuramoto model ~\cite{ref:Kuramoto} consists in a finite population of $N$ globally-coupled phase
oscillators governed by
\begin{equation}
{d\phi_j\over dt}=\omega_j-\frac{K}{N}\sum_{k=1}^{N}\sin(\phi_j-\phi_k),~~j=1,\cdots,N.
\label{eq:model}
\end{equation}
Here $\phi_j$ is the phase of the $j$th oscillator, and $\omega_j$ its intrinsic frequency.
A {\it given realization} of the Kuramoto model corresponds to a particular choice of the $N$
intrinsic frequencies $\{\omega_j\}$. In this paper, we shall assume that the number density of these frequencies
on the real axis follows, on average, the normal distribution $g(\omega)=(1/\sqrt{2\pi})e^{-\omega^2/2}$.
The coupling strength $K$ is positive ($K>0$).

Phase synchronization is conveniently described by the complex order parameter
defined as~\cite{ref:Kuramoto}
\begin{equation}
Z(t) \equiv \Delta e^{i\theta} = \frac{1}{N}\sum_{j=1}^{N} e^{i\phi_j},
\label{eq:def_Delta}
\end{equation}
where a nonzero (positive) value of $\Delta$ in the asymptotic $N\to\infty$ limit
implies the emergence of phase synchronization, and
$\theta$ is the phase of the global order parameter.
With this definition, Eq.~(\ref{eq:model}) is rewritten:
\begin{equation}
\dot{\phi_j} = \omega_j - K\Delta \sin(\phi_j-\theta),
\label{eq:model_withDelta}
\end{equation}
where both $\Delta$ and $\theta$ are dynamic variables defined by Eq. (\ref{eq:def_Delta}).
In the limit $N\rightarrow\infty$, the model exhibits a continuous
symmetry-breaking transition at $K=K_c=2/[\pi g(0)]=\sqrt{8/\pi}$~\cite{ref:Kuramoto}.
In this limit, the phase order parameter is strictly zero on the subcritical side ($K<K_c$), while on
the supercritical side ($K>K_c$),
\begin{equation}
\Delta\sim (K-K_c)^\beta \quad \text{with} \quad \beta=\frac{1}{2}.
\label{eq:Kuramoto}
\end{equation}
This result holds for any unimodal and symmetric $g(\omega)$.

For a finite population of oscillators, $\Delta(t)$ exhibits both dynamic and sample-to-sample
fluctuations whose scaling with $N$ is the main focus of this work.
The sample-to-sample fluctuations arise from the randomly drawn oscillator frequencies from
the distribution $g(\omega)$. Consequently, the spacing between neighboring frequencies
on the real axis is Poissonian in any given realization.
It turns out that this quenched noise and the nonlinear dynamic fluctuations are intimately related to each other
in the sense that, when the former is removed, a very different set of FSS exponents are obtained.
The two types oscillator frequency distributions are discussed in Sec. III and IV, respectively.

\section{Random frequency distribution}

In this section, we consider the realizations (samples) of frequency sets  $\{\omega_j\}$
by randomly and independently  drawing frequencies from the Gaussian distribution $g(\omega)$.
We measure the phase order parameter and its dynamic fluctuation for each sample
and average over many samples. In practice, we numerically integrate Eq.~(\ref{eq:model_withDelta}),
for system sizes up to $N=12\ 800$,
using the Heun's method~\cite{ref:Heun} with a discrete time step $\delta t=0.01$, up to
total $2\times 10^5$ time steps. We start with random initial values of $\{\phi_j(0)\}$ and
average the data over time after the first $10^5$ steps to avoid any transient behavior
and also average over $10^2 - 10^3$ different samples of $\{\omega_j\}$ and $\{\phi_j(0)\}$.
At $K=K_c$, we perform numerical integration for larger systems up to $N= 204\ 800$  with twice more realizations.

\subsection{Finite-size scaling of the order parameter}

In analogy with equilibrium critical phenomena, we expect the order parameter in the critical region
to satisfy the FSS~\cite{ref:UHP},
\begin{equation}
\Delta(K, N)=N^{-\beta/\bar\nu}f(\epsilon N^{1/\bar\nu})~,
\label{eq:Delta_random_FSS}
\end{equation}
where $\epsilon=K-K_c$ and the scaling function $f(x)$ is a monotonically
increasing function with the following asymptotic properties,
\begin{equation}
f(x)  \sim
\left\{
\begin{array}{ll}
x^{\beta}, & x \gg 1;\\
(-x)^{\beta-\bar\nu/2}, & x \ll -1.
\end{array}
\right.
\label{eq:f}
\end{equation}
The exponent $\bar\nu$ is known as the FSS exponent.
For large $N$, this scaling ansatz reproduces Eq. (\ref{eq:Kuramoto})
on the supercritical side, and yields
$\Delta\sim N^{-\beta/{\bar\nu}}$ at the transition and $\Delta \sim N^{-\frac{1}{2}}$ in the subcritical regime,
respectively.

In the usual mean-field treatment of the Kuramoto model, the entrained state
is established by solving Eq.~(\ref{eq:model_withDelta})
at a constant $\Delta$ and $\theta$ (dynamic fluctuations are ignored):
Oscillators with $|\omega_j|< K\Delta$ get into an entrained (locked) state, reaching
a fixed angle. On the other hand, the oscillators with $|\omega_j|> K\Delta$
perform periodic motion with modified frequencies.
With this, the self-consistent equation for $\Delta$ from Eq.~(\ref{eq:def_Delta}), can be written as
\begin{equation}
\Delta = \int_{-K\Delta}^{K\Delta} d\omega g(\omega) \sqrt{1-(\omega/K\Delta)^2}
\label{eq:Psi_Delta}
\end{equation}
in the $N\rightarrow \infty$ limit.  Eq.~(\ref{eq:Psi_Delta}) has a
nontrivial solution ($\Delta>0$) for $K>K_c=2/[\pi g(0)]$.

For a finite-size system, the right-hand side of Eq.~(\ref{eq:Psi_Delta}) is replaced with
\begin{equation}
\Delta=\Psi(K\Delta) \equiv \frac{1}{N}\sum_{j, |\omega_j|<K\Delta}\sqrt{1-(\omega_j/K\Delta)^2},
\label{eq:SCE-FS}
\end{equation}
which contains a sample-dependent correction
$\delta\Psi\equiv \Psi-\left[ \Psi \right] \propto \sqrt{N_s}/N$~\cite{ref:HCPT_PRL}. Here
$N_s$ is the number of (entrained) oscillators in the frequency interval
$(-K\Delta, K\Delta)$ and $\left[\cdot\right]$ denotes sample average.
As was shown previously in Refs.~\cite{ref:HCPT_PRL,ref:Tang}
and explained in detail in the Appendix A, for $K$ close to $K_c$,
Eq.~(\ref{eq:SCE-FS}) admits a scaling solution of the form (\ref{eq:Delta_random_FSS})
with the exponents
\begin{equation}
\beta=\frac{1}{2}, \quad \text{and}\quad \bar\nu=\frac{5}{2}~.
\end{equation}
The scaling function $f(x)$, however, has significant sample-to-sample fluctuations.

The exponent value $\beta=\frac{1}{2}$ is typical of various equilibrium mean-field models
described by the so-called $\phi^4$ theory. However, the FSS exponent value of $\bar\nu=\frac{5}{2}$ is quite unusual
and implies much stronger fluctuations than that in ordinary models with the conventional value $\bar\nu=2$.
This remarkable result has been confirmed by extensive numerical simulations~\cite{ref:HPC_PRE,ref:HCPT_PRL}
and the straightforward extension to the Kuramoto model on various sparse random networks
 in~\cite{ref:HPT_Synch_SFN,ref:HUP,ref:UHP}.

\subsection{Dynamic fluctuations}

Now, we consider the dynamic (temporal) fluctuations of
the order parameter by measuring the quantity~\cite{ref:Daido-ex}
\begin{equation}
\chi (K, N)\equiv N\left[\langle (\Delta - \langle\Delta\rangle)^2\rangle\right] =N\left[\langle \Delta^2\rangle \!-\! \langle \Delta\rangle^2\right]
\label{eq:chi_def}
\end{equation}
where $\langle \cdot \rangle$ denotes time average in the steady state of a given sample and
$\left[\cdot\right]$ sample average, respectively.

For  $K<K_c$, we expect from simple power counting that $\chi \sim {\cal O}(1)$, because
$\Delta\sim N^{-\frac{1}{2}}$. For $K>K_c$, the order parameter becomes ${\cal O}(1)$, but
these ${\cal O} (1)$ terms cancel out each other exactly and the next order terms proportional to $N^{-\frac{1}{2}}$ contribute.
Again, $\chi$ becomes ${\cal O}(1)$.
However, at $K=K_c$, strong critical FSS occurs since $\Delta\sim N^{-\beta/\bar\nu}=N^{-\frac{1}{5}}$,
and one can expect that $\chi$ will diverge in the $N\rightarrow\infty$ limit.
Summing up, the critical behavior of $\chi$ near the transition may be expressed,  in the $N \rightarrow \infty$ limit, as
\begin{equation}
\chi  \sim
\left\{
\begin{array}{ll}
\epsilon^{-\gamma}, &  \epsilon>0; \\
(-\epsilon)^{-\gamma^{\prime}}, & \epsilon<0.
\end{array}
\right.
\label{eq:chi_gamma}
\end{equation}
The two exponents $\gamma$ and $\gamma^{\prime}$ describe the divergence of
$\chi$ in the supercritical~($\epsilon>0$) and subcritical~($\epsilon<0$) regions, respectively.

For homogeneous systems, scaling in the critical region is controlled by a single
fixed point in the renormalization group (RG) sense, and hence
$\gamma=\gamma'$. The corresponding FSS takes the form
\begin{equation}
\chi(\epsilon, N)=N^{\gamma/\bar\nu}h(\epsilon N^{1/\bar\nu})~,
\label{eq:FSS}
\end{equation}
irrespective of the side of the transition,
and the scaling function $h(x)$ has the limiting behavior
\begin{equation}
h(x)  \sim
\left\{
\begin{array}{ll}
{\mbox {const}}, & x=0; \\
x^{-\gamma}, & x \gg 1; \\
(-x)^{-\gamma}, & x \ll -1.
\end{array}
\right.
\label{eq:scalingfunction}
\end{equation}

In this context, the Kuramoto model is somewhat special in that each
oscillator has its own intrinsic frequency and the effect of the
mean-field is different on different oscillators. Therefore the usual RG argument with only
a few relevant variables may not apply
and one should entertain the possibility that $\gamma\neq \gamma'$. In such a situation,
the FSS exponent $\bar\nu$ should also be different on the two sides of the transition,
so that the ratio relation $\gamma/\bar\nu=\gamma'/\bar\nu'$ be satisfied,
in order to maintain the continuity of $\chi$ at $\epsilon=0$.
Indeed, through perturbative calculation, Daido~\cite{ref:Daido} obtained $\gamma=\frac{1}{4}$ and $\gamma'=1$.

\begin{figure}
\epsfxsize=\linewidth \epsfbox{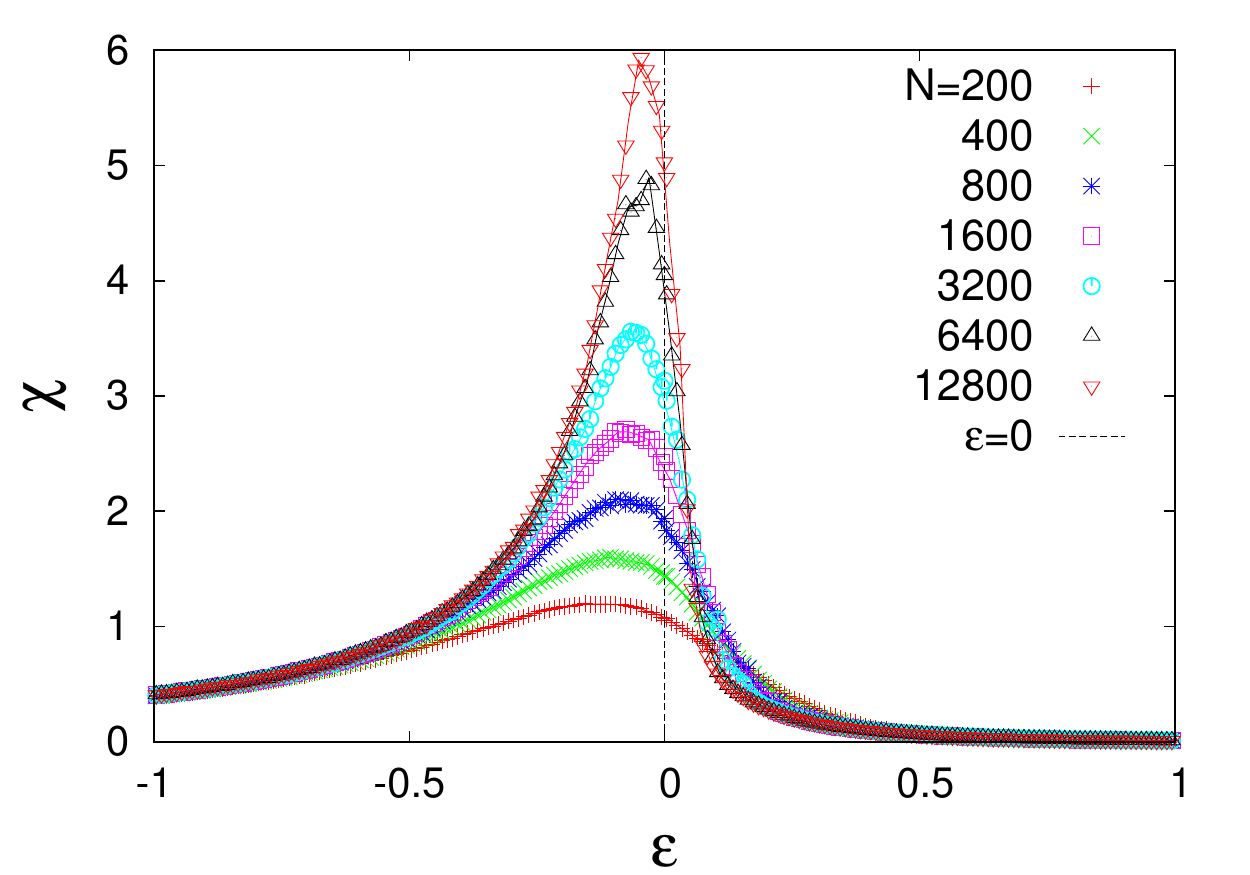}
\caption{(Color online) Dynamic fluctuation $\chi$ against
$\epsilon=K-K_c$ for various system size $N$, with $K_c=\sqrt{8/\pi}$.
The data have been obtained from the average over $100 - 1000$ samples with different
sets of $\{\omega_j \}$ and initial phases $\{\phi_j(0)\}$. Statistical errors
are represented by symbol sizes.
}
\label{fig:chi_random}
\end{figure}

\begin{figure}
\epsfxsize=\linewidth \epsfbox{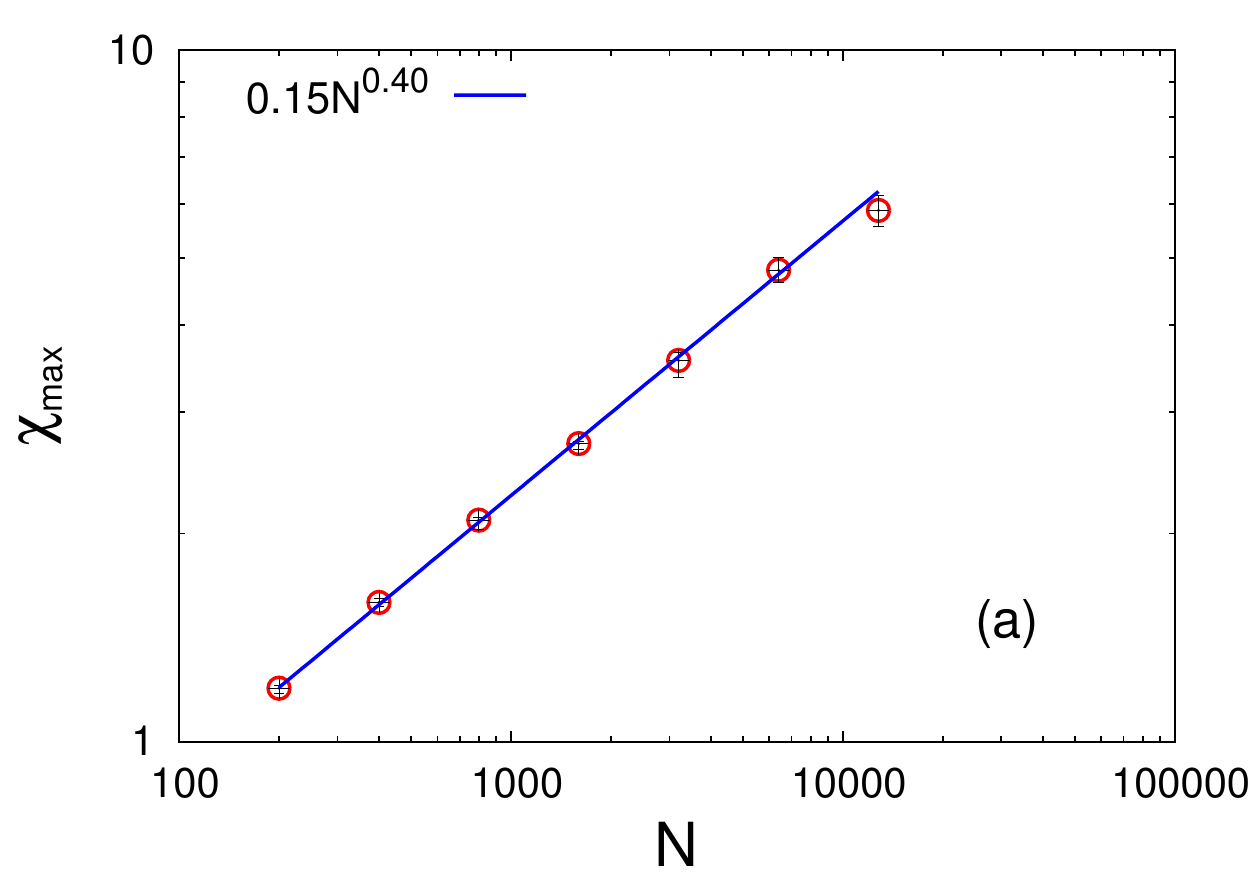}
\epsfxsize=\linewidth \epsfbox{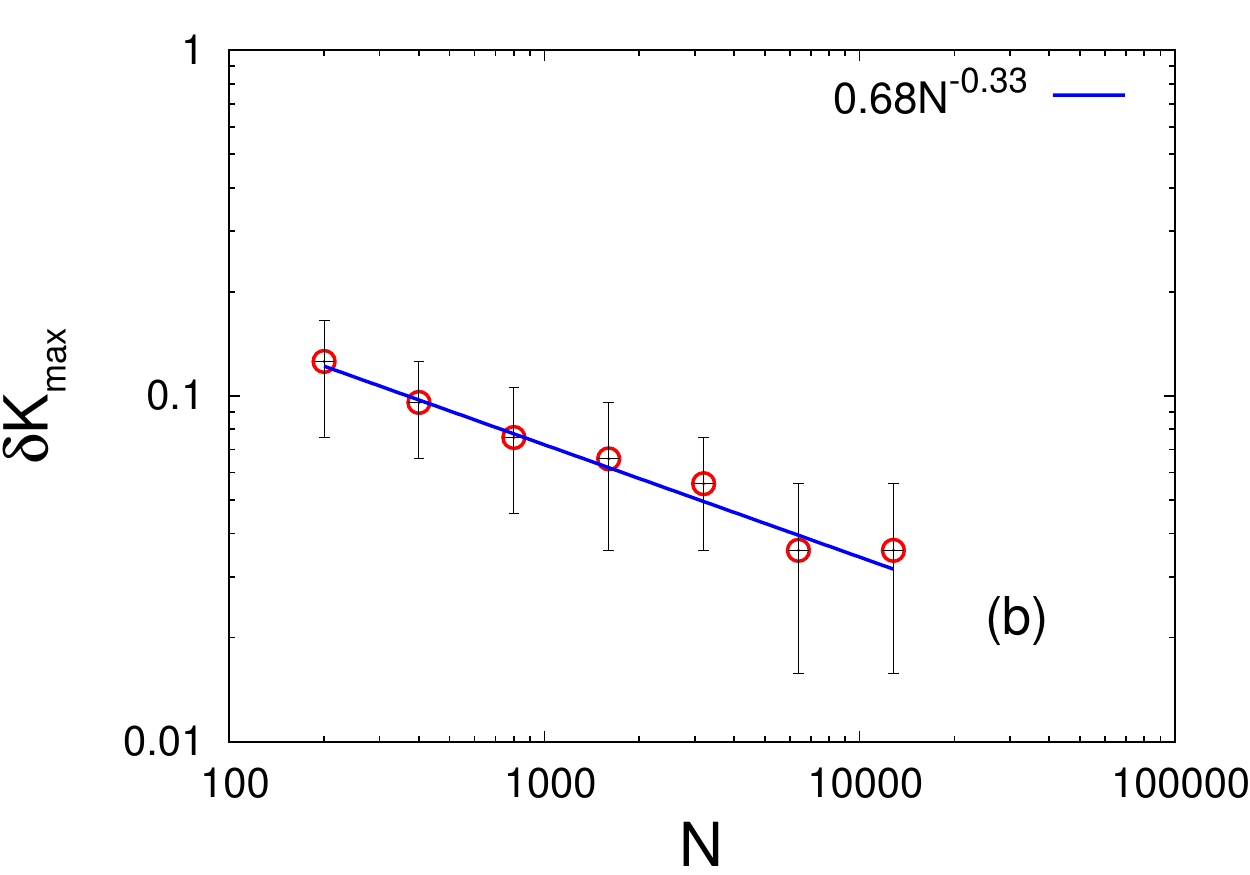}
\caption{(Color online)
(a) Behavior of $\chi_{\rm max}$ is shown as a function of $N$ in a log-log plot.
The straight line obtained through the least-square fitting method displays $\chi_{\rm{max}}\sim N^{0.40}$.
(b) Behavior of $\delta K_{\rm{max}}(=|K_{\rm{max}} - K_c|)$ is shown as a function of $N$ in a log-log plot.
The straight line obtained through the least-square fitting method displays $\delta K_{\rm{max}} \sim N^{-0.33}$.
}
\label{fig:chimax_random}
\end{figure}

\begin{figure}
\epsfxsize=\linewidth \epsfbox{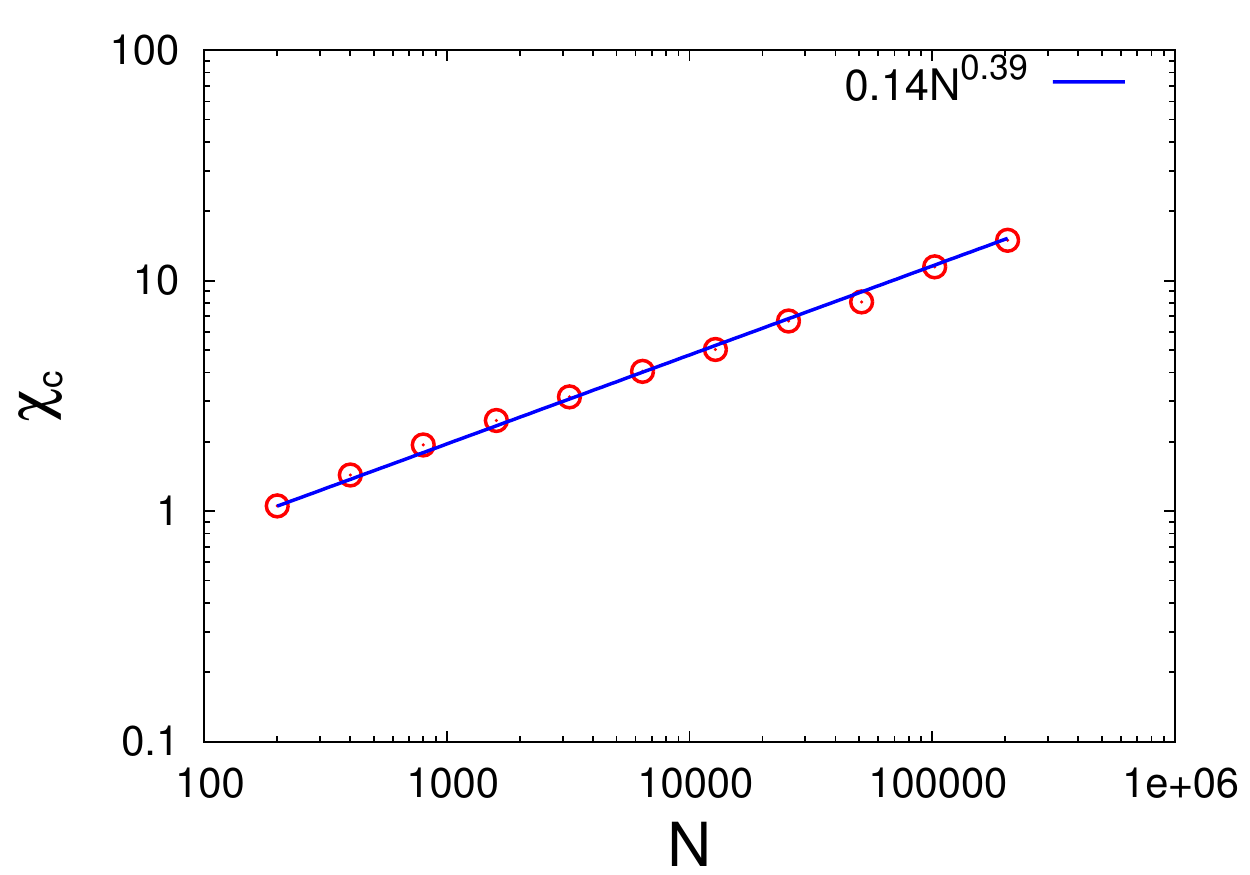}
\caption{(Color online)
Critical increase of $\chi$ at $K_c(=\sqrt{8/\pi})$ is plotted as a function of $N$ in a log-log plot.
The slope of the straight line is 0.39. Statistical errors are represented by symbol sizes.
}
\label{fig:chiatKc_random}
\end{figure}

We have carried out extensive numerical simulations to obtain accurate estimates
of $\gamma$ and $\gamma'$ as well as other FSS exponents.
Figure~\ref{fig:chi_random} shows numerical data for  $\chi$
versus $\epsilon$ for various system sizes $N$.
As expected, finite-size effects become huge near the transition point.
We observe two important features of the finite-size effects: First, the peak position at $K=K_{\rm{max}}$ is always
on the subcritical side ($\epsilon<0$)
and approaches the bulk critical point ($\epsilon=0$) as $N$ increases. Second, the peak height
$\chi_{\rm{max}}$ grows rapidly with $N$. By analyzing the data for the peak position and height, we
estimate $\gamma'$ and $\bar\nu'$ in the subcritical side, using the conventional FSS ansatz
\begin{equation}
\label{eq:chimax_form} \chi_{\rm{max}}\sim N^{\gamma^{\prime}/{{\bar\nu}^{\prime}}}, \qquad
\delta K_{\rm{max}}  \sim N^{-1/{{\bar\nu}^{\prime}}}
\end{equation}
with $\delta K_{\rm{max}}\equiv |K_{\rm{max}}-K_c|$.
As shown in Fig.~\ref{fig:chimax_random}~(a,b), both quantities exhibit
power-law scaling with the exponent values
\begin{equation}
\label{eq:sub_exponents} {\gamma^{\prime}/{{\bar\nu}^{\prime}}}= 0.40(1), \qquad
1/{{\bar\nu}^{\prime}}= 0.33(8) ~.
\end{equation}
Even though our estimate of $1/\bar\nu'$ is rather weak due to huge statistical uncertainties in locating
$K_{\rm{max}}$, it includes our analytic value of $1/\bar\nu=\frac{2}{5}$ within the error bar.
Taking a simple ratio of the above exponent values, we estimate $\gamma'=1.2(3)$. If we take
$1/\bar\nu'=\frac{2}{5}$, then  $\gamma'=1.00(3)$.

We also measure the value of $\chi$ at the bulk critical point $K=K_c$. As $K_c$ is located to the right of the peak position $K_{\rm{max}} (N)$  for all $N$
in this case, the FSS behavior at $K_c$ may be regarded as a continuation of the scaling on the supercritical side.
Again, the FSS ansatz leads to
\begin{equation}
\chi_c \equiv \chi(\epsilon=0, N) \sim N^{\gamma/\bar\nu} ~.
\label{eq:chiatKc_form}
\end{equation}
Figure~\ref{fig:chiatKc_random} shows that the numerical data agree almost perfectly with
the FSS ansatz with
\begin{equation}
\label{eq:super_exponents} {\gamma/{{\bar\nu}}}= 0.39(2) ~.
\end{equation}
This result confirms the ratio relation $\gamma/\bar\nu=\gamma'/\bar\nu'$,
but does not provide estimates for $\gamma$ and $\bar\nu$ separately.

\begin{figure}
\epsfxsize=\linewidth \epsfbox{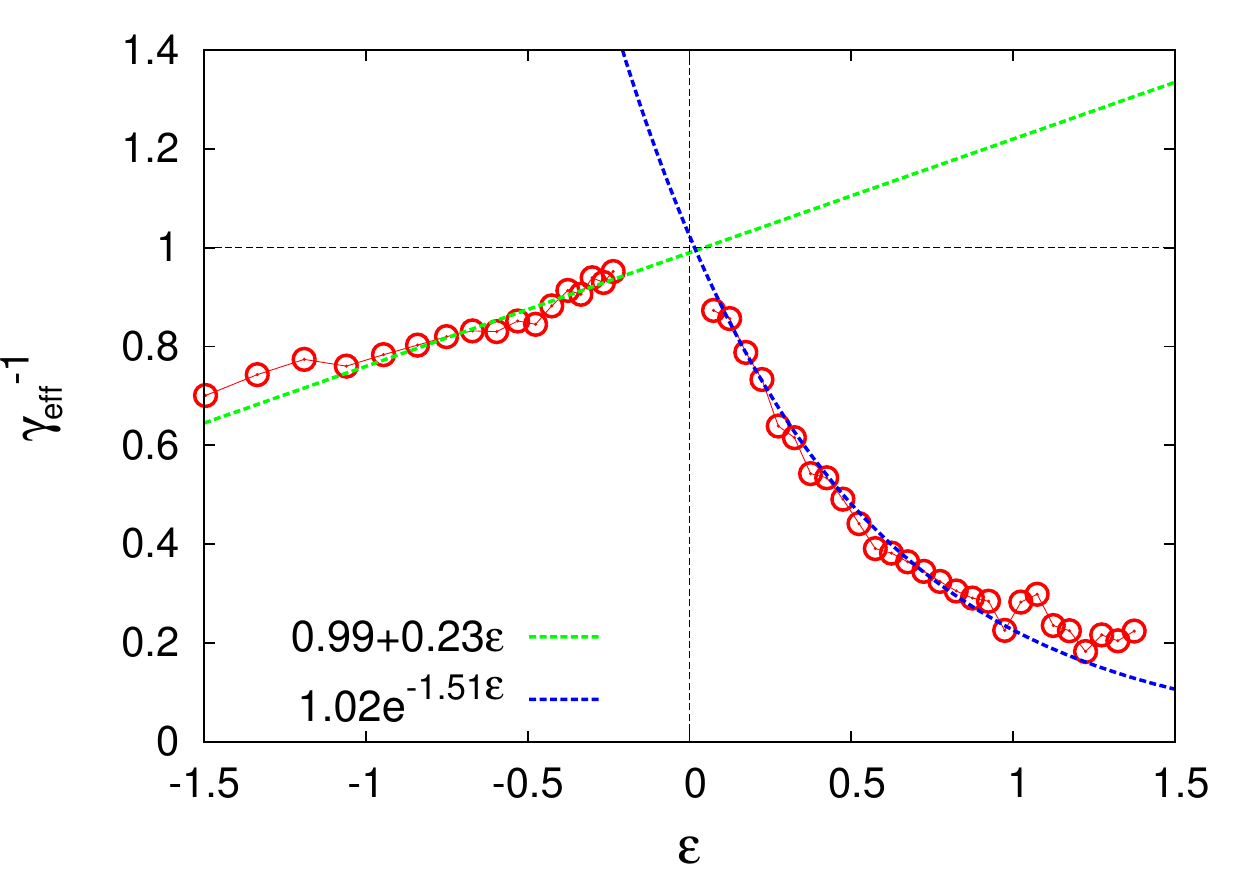}
\caption{(Color online)
Numerical data for the inverse of the effective exponent
${\gamma^{-1}_{\rm{eff}}}$ are plotted against $\epsilon$ for $N= 12\ 800$.
Two dotted  lines represent the fitting lines,
indicating $\gamma\approx 0.98$ and $\gamma' \approx 1.01$.}
\label{fig:gammaeff_random}
\end{figure}

To estimate $\gamma$ and $\gamma'$ directly from off-critical data in Eq.~(\ref{eq:chi_gamma}),
we examine the local slopes of $\chi$ against $|\epsilon|$ in double logarithmic plots
 for very large $N$. The {\em effective} exponent $\gamma_{\rm{eff}}$ is defined as
\begin{equation}
\gamma_{\rm{eff}}(\epsilon)\equiv \frac{d \ln \chi}{d \ln|\epsilon|},
\label{eq:def_gammaeff}
\end{equation}
and the critical exponents are obtained from their asymptotic values as
$\gamma =\lim_{\epsilon\rightarrow 0^+}\gamma_{\rm{eff}}(\epsilon)$ and $\gamma' =\lim_{\epsilon\rightarrow 0^-}\gamma_{\rm{eff}}(\epsilon)$.
Figure~\ref{fig:gammaeff_random} shows the inverse of the effective
exponent $\gamma^{-1}_{\rm eff}$
as a function of $\epsilon$.
We find that $\gamma_{\rm{eff}}^{-1}$ approaches $\gamma'^{-1}$ linearly
as $\epsilon$ goes to zero from below in the subcritical region,
with its asymptotic value $\gamma'=1.01(3)$. This is consistent with our previous numerical estimates
and also with the theoretical prediction of $\gamma'=1$ by Daido~\cite{ref:Daido}.
Note that $\gamma'=1$ has been also obtained in a systematic $1/N$ expansion~\cite{ref:Hildebrand},
where the random frequency distribution is prerequisite.

On the supercritical side, $\gamma_{\rm{eff}}^{-1}$ varies in a nonlinear fashion
far from the transition, but  approaches $\gamma^{-1}$ linearly  near small $\epsilon$
with the asymptotic value $\gamma=0.98(5)$. This suggests strongly that $\gamma=\gamma'=1$, and
definitely excludes Daido's value $\gamma=\frac{1}{4}$. With $\gamma=1$, our numerical estimate
in Eq.~(\ref{eq:super_exponents}) lead to $\bar\nu=\bar\nu'=\frac{5}{2}$.

Summing up all these results, we conclude that our numerical data strongly support simple scaling exponents:
\begin{equation}
\beta=\frac{1}{2}, \quad \bar\nu=\bar\nu'=\frac{5}{2},  \quad \text{and} \quad \gamma=\gamma'=1 ~.
\label{eq:conjecture_random}
\end{equation}
However, a proper analytic treatment for the dynamic fluctuations in the supercritical side
is lacking, and further investigations are thus needed.
One should note that the dynamic fluctuations of the order parameter are not necessarily proportional to the susceptibility
(response of the order parameter to an infinitesimal external field) in general nonequilibrium steady states
due to probable violation of the fluctuation-dissipation theorem.
It is almost trivial to derive the susceptibility
exponents in the Kuramoto model, which turn out to be $\gamma_{\rm{sus}}=\gamma'_{\rm{sus}}=1$~\cite{ref:Daido_PRE}.
Yet this accordance with the dynamic fluctuation exponents
in Eq.~(\ref{eq:conjecture_random}) may be merely coincidental.

\subsection{Hyperscaling relation}

In equilibrium critical phenomenon, the equation
\begin{equation}
\gamma=\bar\nu-2\beta ~,
\label{eq:hsr}
\end{equation}
with $\bar\nu = d\nu$, is usually referred to as a hyperscaling relation.
Here $d$ is the dimension of the system and $\nu$ is the correlation length exponent.
Equation (\ref{eq:hsr}) follows from a simple power counting of Eq.~(\ref{eq:chi_def}) at the transition.
The left-hand side of  Eq.~(\ref{eq:chi_def}) is $\chi_c\sim N^{\gamma/\bar\nu}$, while
the right-hand side is proportional to $N^{1-2\beta/\bar\nu}$ if the leading order terms do not exactly
cancel out each other.

\begin{figure}
\epsfxsize=\linewidth\epsfbox{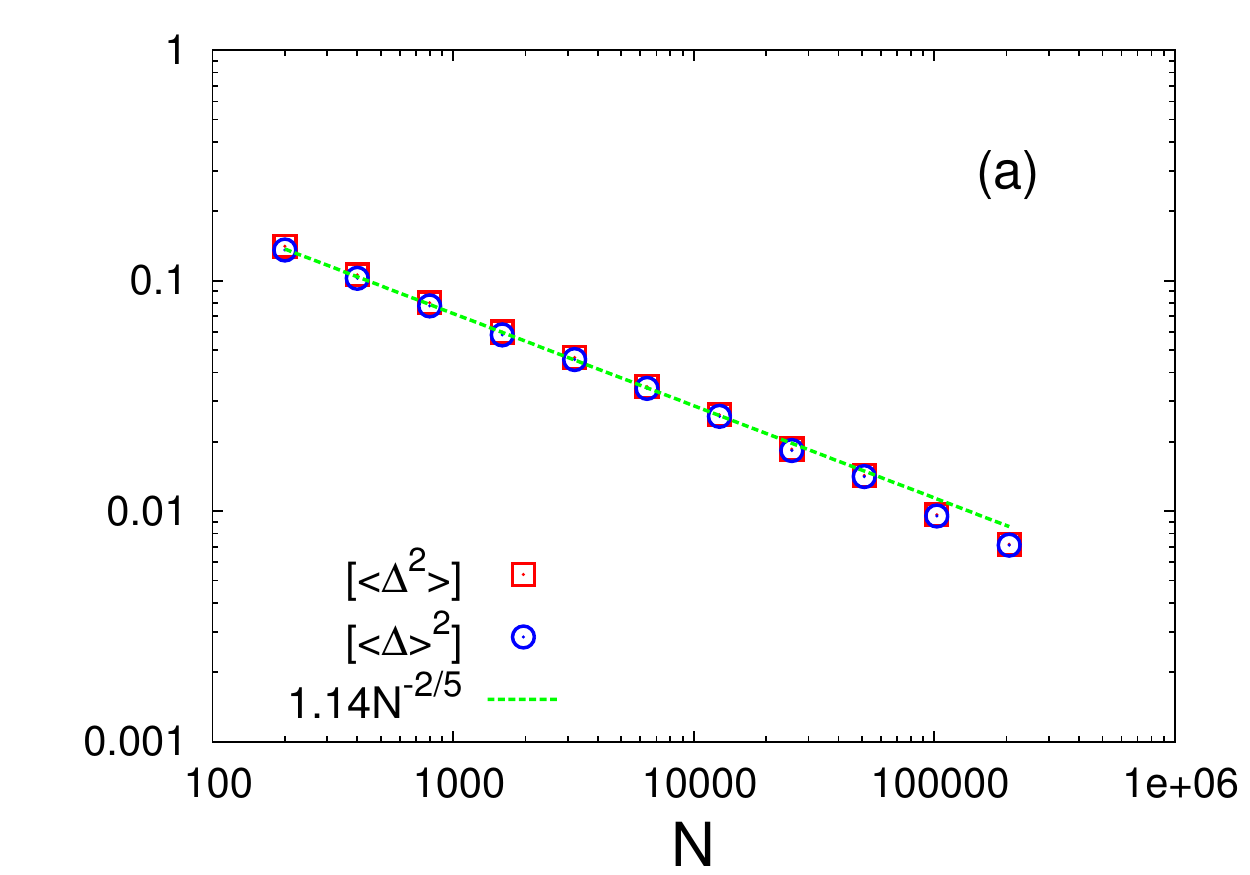}
\epsfxsize=\linewidth\epsfbox{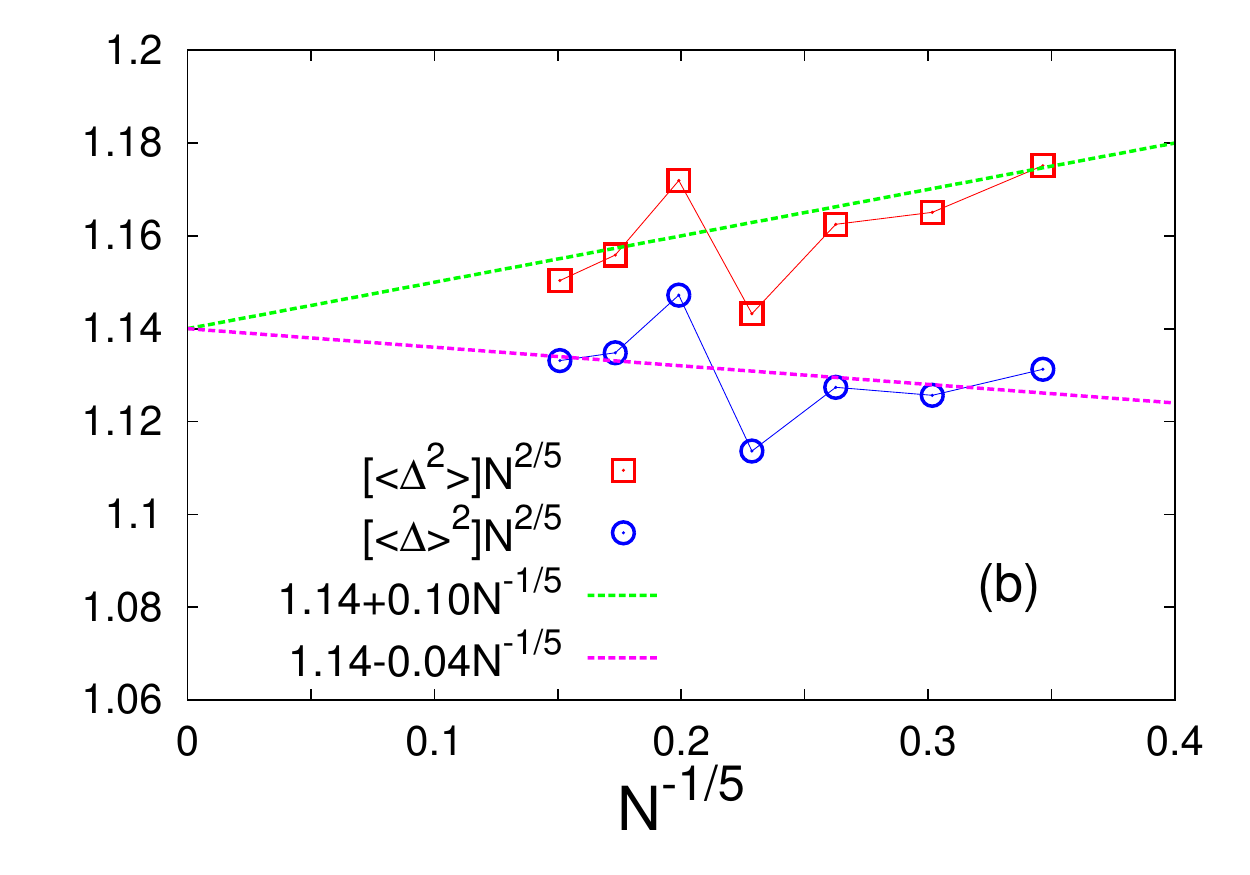}
\caption{(Color online)
(a) Critical decay of $\left[\langle\Delta^2\rangle\right]$ and $\left[\langle\Delta\rangle^2\right]$ at
$K_c(=\sqrt{8/\pi})$ are plotted against $N$ in a log-log plot. These two data are so close to each other that
it is not possible to discern two data in this plot. The straight line is the fitting line of $1.14 N^{-0.40}$.
(b) $N^{\frac{2}{5}}\left[\langle\Delta^2\rangle\right]$ and $N^{\frac{2}{5}}\left[\langle\Delta\rangle^2\right]$ versus $N^{-\frac{1}{5}}$.
These data are expected to show straight lines. The least-square fitting yields two straight lines with the same intercept.
Lines between data are only guides to the eye.}
\label{fig:chi1chi2_random}
\end{figure}

It is obvious that our result Eq.~(\ref{eq:conjecture_random}) does not satisfy this relation.
Violating the hyperscaling relation is only possible when the leading order terms in $\left[\langle \Delta^2\rangle\right]$ and
$\left[\langle \Delta\rangle^2\right]$
cancel out exactly and the sub-leading order terms show up, yielding $\gamma=1$
instead of $\frac{3}{2}$. A plausible conjecture for leading and sub-leading terms for the large $N$ expansion of the
two quantities is then
\begin{eqnarray}
\left[\langle\Delta^2\rangle\right] &=& N^{-2\beta/\bar\nu} \left(c + d N^{-\frac{1}{5}}\right)~, \nonumber\\
\left[\langle\Delta\rangle^2\right] &=&  N^{-2\beta/\bar\nu}\left(c'  + e N^{-\frac{1}{5}}\right)~,
\label{eq:Delta_square}
\end{eqnarray}
with $c=c'$.
Indeed, our numerical data show in Fig.~\ref{fig:chi1chi2_random}  that $c\approx c'=1.140(2)$,
$d=0.10(4)$, and $e=-0.04(4)$.
Then, the dynamic fluctuation $\chi$ at the transition scales as
\begin{equation}
\chi_c = (d-e) N^{1-2\beta/\bar\nu-\frac{1}{5}}\sim N^{\gamma/\bar\nu} ~,
\end{equation}
yielding
\begin{equation}
\gamma=\bar\nu-2\beta-\frac{1}{2}=1.
\end{equation}

The cancellation of the leading order ($c=c'$) in Eq.~(\ref{eq:Delta_square}) would be expected if dynamic fluctuations only
produce subleading order corrections to the solution of the mean-field equation (\ref{eq:SCE-FS}) at the transition.
Interestingly, as we show in Appendix A, the mean-field approximation yields
$[\langle\Delta^2\rangle]=[\langle\Delta\rangle^2]\simeq 0.93 N^{-\frac{2}{5}}$ which is smaller in amplitude
as compared to the numerically determined value $c=1.14$. Therefore it appears that dynamic fluctuations
renormalize the strength of the quenched noise $\delta\Psi$, but
do not change its scaling with $N$.

It is interesting to compare the violation of hyperscaling in the Kuramoto model to known cases in equilibrium phase
transitions. One well-known such case is the incompatibility between the
correlation length exponent $\nu=\frac{1}{2}$ at the Gaussian fixed point and the mean-field exponents $\beta=\frac{1}{2}$
and $\gamma=1$ above the critical dimension $d_c=4$~\cite{ref:HHP}. Violation of hyperscaling is also reported in the
three-dimensional random-field Ising model whose ordering transition is controlled by a zero-temperature
fixed point\cite{ref:Villain,ref:Fisher,ref:Nattermann,ref:Binder}. In both cases, the energy associated with the
symmetry-breaking field is much stronger than the thermal energy on large length scales, hence the order
parameter fluctuations, as measured by the exponent $\gamma$, become less than what is predicted
by the hyperscaling relation. In this sense, the weaker dynamic fluctuations of the order parameter in the Kuramoto
model may be interpreted in a similar way, despite the fact that we are now in a nonequilibrium situation.

\section{Regular frequency distribution}

In this section, we consider the `regular' frequency set $\{\omega_j\}$ with minimal disorder among
frequencies. This set can be generated following a deterministic procedure given by
\begin{equation}
\frac{j-0.5}{N}=\int_{-\infty}^{\omega_j} g(\omega)d\omega,
\label{eq:w_determin}
\end{equation}
with $j=1,\ldots, N$ and $g(\omega)=(1/\sqrt{2\pi})e^{-\omega^2/2}$.
The generated frequencies are quasi-uniformly spaced in accordance with $g(\omega)$ in the population.
As this set is uniquely determined for each $N$, sample disorder may result from initial conditions only.
Note that the regular frequency distribution is considered in numerical attempts by Daido~\cite{ref:Daido}.

Similar to the case of the random distribution, we numerically integrate Eq.~(\ref{eq:model_withDelta})
for system sizes up to $N=12\ 800$ with a discrete time step $\delta t=0.01$, up to
total $10^7$ time steps. We start with random initial values of $\{\phi_j(0)\}$ and
average the data over the latter half of time steps and also average over $20 - 100$ different samples of $\{\phi_j(0)\}$.
At $K=K_c$, we perform numerical integration for larger systems up to $N= 25\ 600$  with $10-100$ samples.

\subsection{Finite-size scaling of the order parameter}

\begin{figure}
\epsfxsize=\linewidth\epsfbox{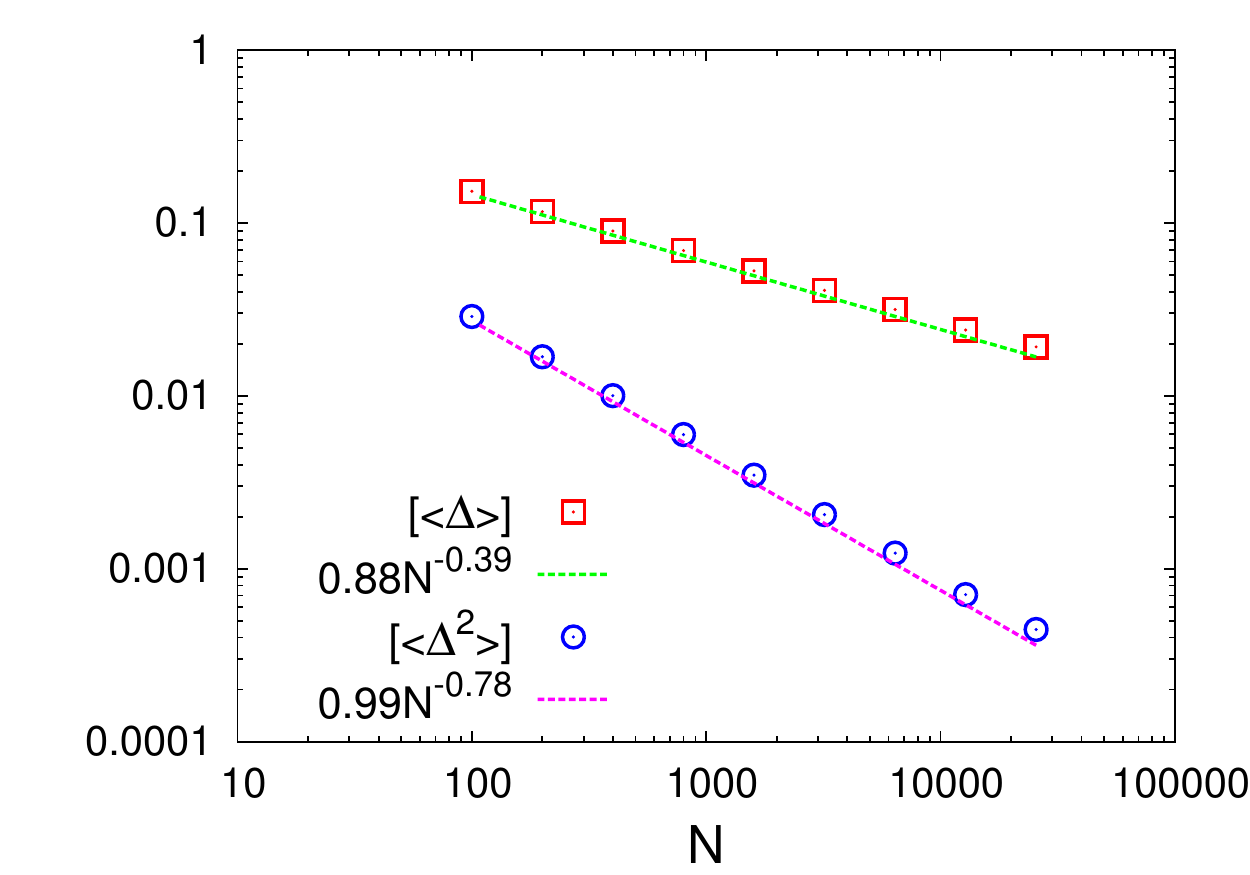}
\caption{(Color online) Critical decay of the order parameter
$\left[\langle\Delta\rangle\right]$ and $\left[\langle\Delta^2\rangle\right]$ at $K_c (=\sqrt{8/\pi})$ is plotted as a
function of $N$ in a log-log plot, which are represented by (red) open boxes and (blue) open circles, respectively.
The slopes of two straight lines are $-0.39$ and $-0.78$, respectively.
}
\label{fig:Delta_N_flessGauss}
\end{figure}

We measure the order parameter and its square during numerical integrations and average over time in the steady state,
and then average over samples with different initial values for the phase variables. These averages are denoted by
$\left[\langle \Delta\rangle\right]$ and $\left[\langle \Delta^2\rangle\right]$, respectively.
Figure~\ref{fig:Delta_N_flessGauss} shows their size dependence at the transition ($\epsilon=0$),
exhibiting the nice power laws
\begin{equation}
\left[\langle \Delta\rangle\right] \sim N^{-0.39(2)} \quad\text{and}\quad \left[\langle \Delta^2\rangle\right] \sim N^{-0.78(3)}~.
\end{equation}
Following the  FSS theory in Eqs.~(\ref{eq:Delta_random_FSS}) and (\ref{eq:f}),
we estimate the decay exponent as
\begin{equation}
\beta/\bar\nu=0.39(2)~.
\end{equation}
Given the value $\beta=\frac{1}{2}$, we obtain
$\bar\nu\approx \frac{5}{4}~.$

\begin{figure}
\epsfxsize=\linewidth \epsfbox{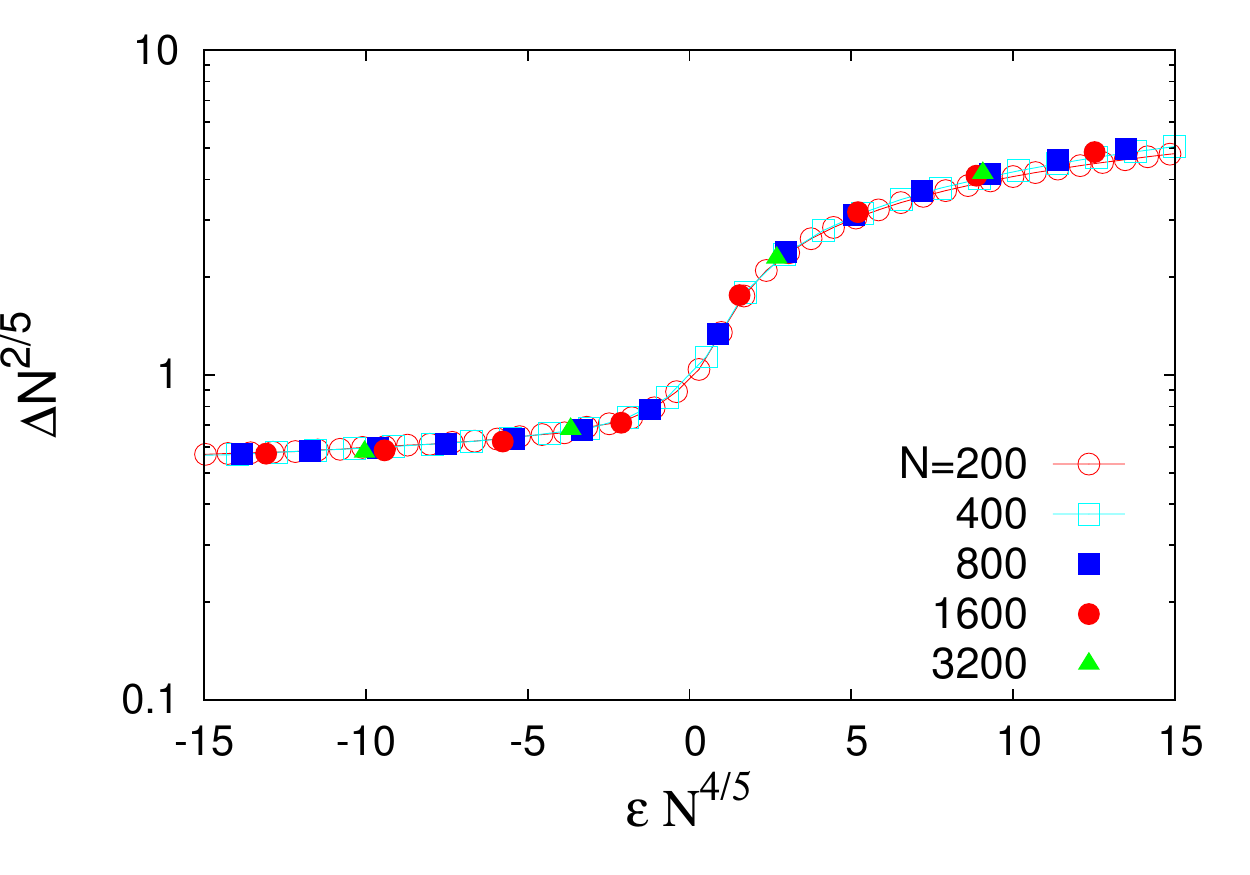}
\caption{(Color online) Scaling plot of the order parameter $\Delta$ for various system size $N$ for the
regular distribution. We use $\beta/\bar\nu=\frac{2}{5}$ and $\bar\nu=\frac{5}{4}$.
}
\label{fig:scaling_flessGauss}
\end{figure}

As a consistency check, we investigate the FSS relation  in Eq.~(\ref{eq:Delta_random_FSS}) for $\epsilon\neq 0$.
We plot $\Delta N^{\beta/\bar\nu}$ versus $\epsilon N^{1/\bar\nu}$ for various system sizes $N$
 in Fig.~\ref{fig:scaling_flessGauss}. (For a moment, we drop the average brackets as
$\Delta \equiv \left[\langle \Delta \rangle\right]$ for convenience.) The data show a perfect collapse
with the choice of $\bar\nu=\frac{5}{4}$. We also considered the regular Lorentzian distribution given by
$g(\omega)=\gamma_{\omega}/[\pi(\omega^2+{\gamma_{\omega}}^2)]$ with the half-width $\gamma_{\omega}$,
where the oscillator frequency is chosen as
$\omega_j =\gamma_{\omega} \tan [j\pi/N-(N+1)\pi/(2N)]$ for $j=1,\ldots,N$.
We find a similar behavior again with $\bar\nu\approx \frac{5}{4}$.

\subsection{Dynamic fluctuations}

\begin{figure}
\epsfxsize=\linewidth \epsfbox{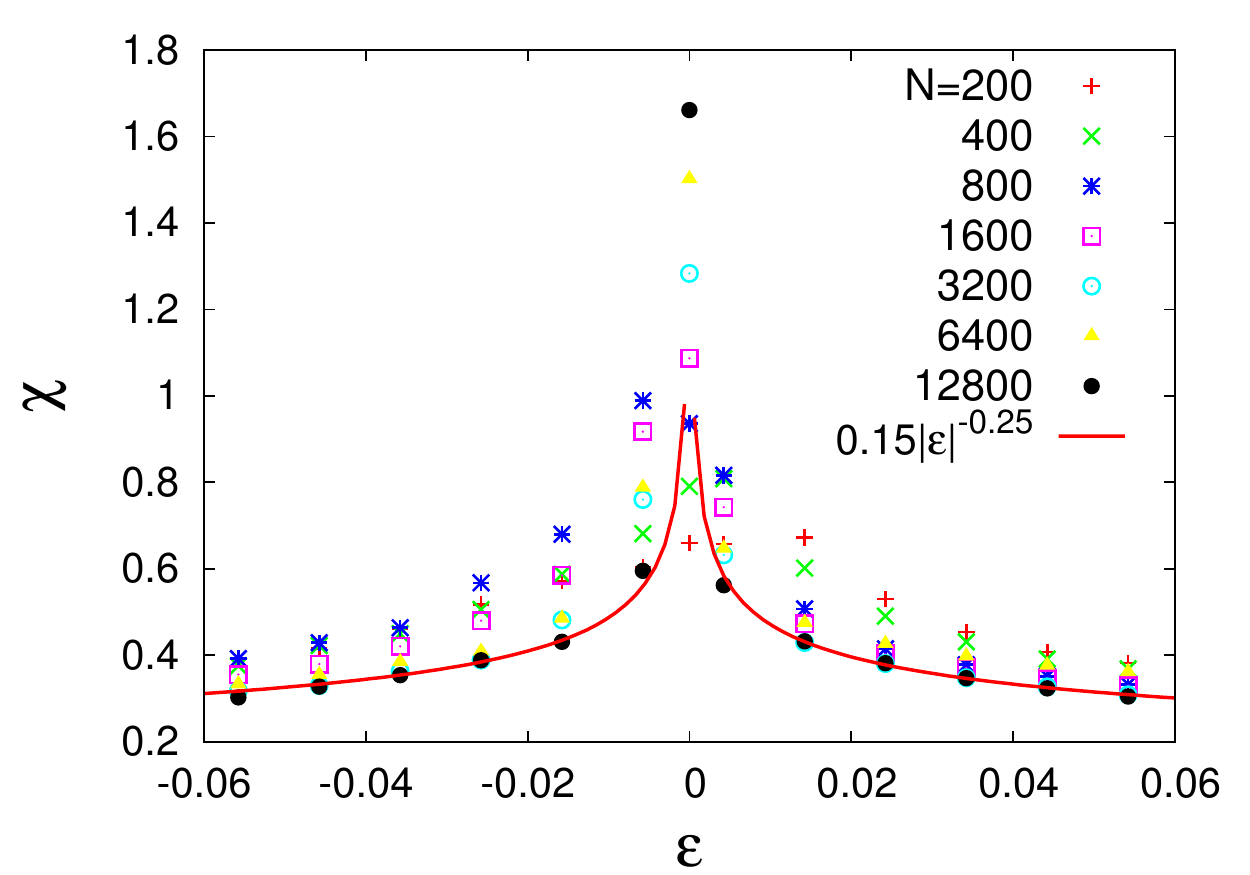}
\caption{(Color online)
Dynamic fluctuation $\chi$ is plotted as a function of $\epsilon(=K-K_c)$ for
various system size $N$ for the regular distribution, where $K_c=\sqrt{8/\pi}$.
The solid (red) line represents a fitting line for data with largest system size.
}
\label{fig:chi_regular}
\end{figure}

\begin{figure}
\epsfxsize=\linewidth \epsfbox{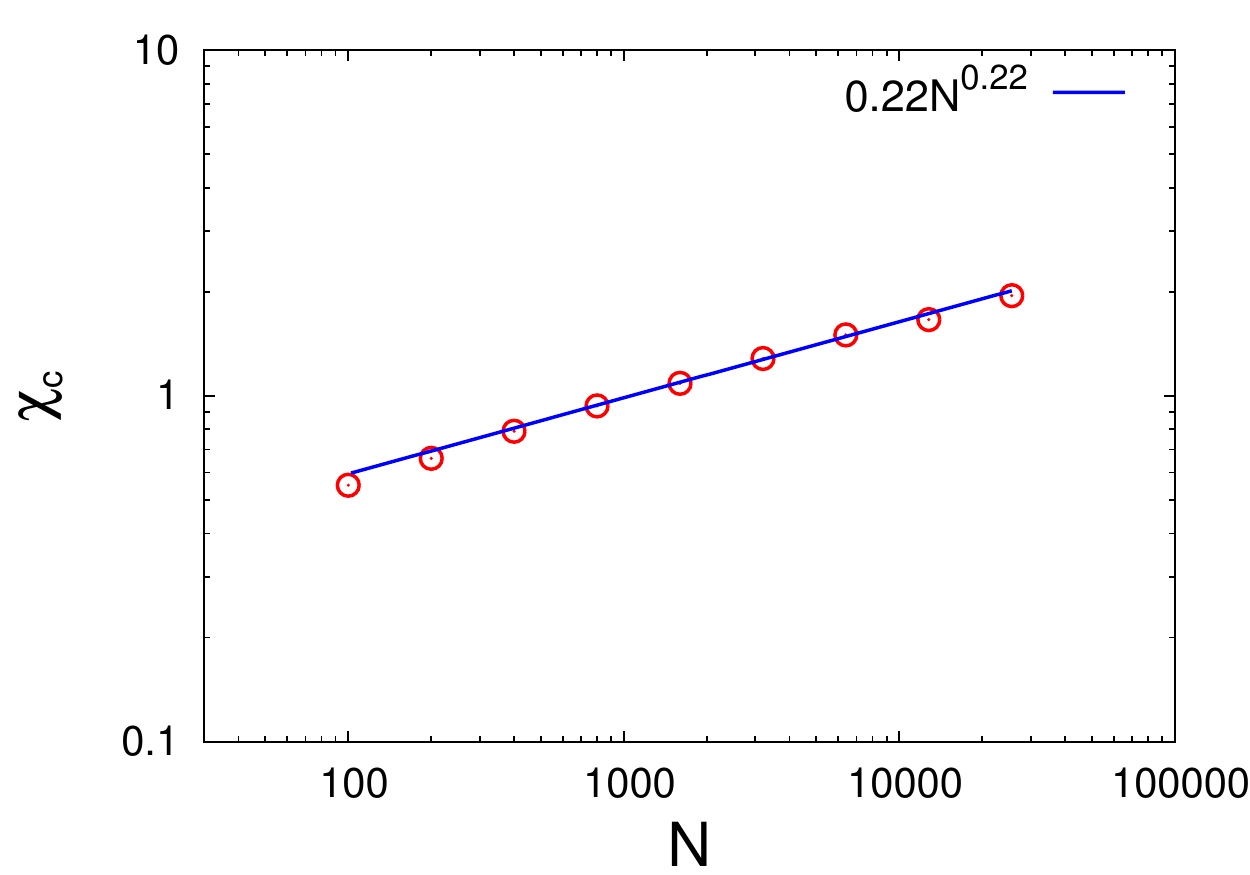}
\caption{(Color online)
Critical increase of $\chi_c$ at $K_c(=\sqrt{8/\pi})$ is plotted as a function of $N$ in a log-log plot for the regular distribution.
The slope of the straight line is $0.22$.
}
\label{fig:chi_atKc_regular}
\end{figure}

In this subsection, we present numerical data for the dynamic fluctuations $\chi$ in Fig.~{\ref{fig:chi_regular}}.
Compared to the random distribution case, $\chi$ is very much reduced in magnitude and peaks only in a much
narrower region around the transition.
The small amplitude makes it difficult to determine the peak height and position accurately.
However, with the exact information of $K_c=\sqrt{8/\pi}$, it is numerically feasible to
investigate the critical increase of $\chi_c$ at the transition as a function of $N$.
The data shown in Fig.~\ref{fig:chi_atKc_regular} fit well to a power law
\begin{equation}
\chi_c\sim N^{\gamma/\bar\nu} \quad \text{with} \quad \gamma/\bar\nu=0.22(2)~.
\end{equation}
Using $\bar\nu=\frac{5}{4}$, we find $\gamma=0.27(3)$, which is close to $\frac{1}{4}$.
This value is consistent with Daido's theoretical result of $\gamma=\frac{1}{4}$~\cite{ref:Daido}.

\begin{figure}
\epsfxsize=\linewidth \epsfbox{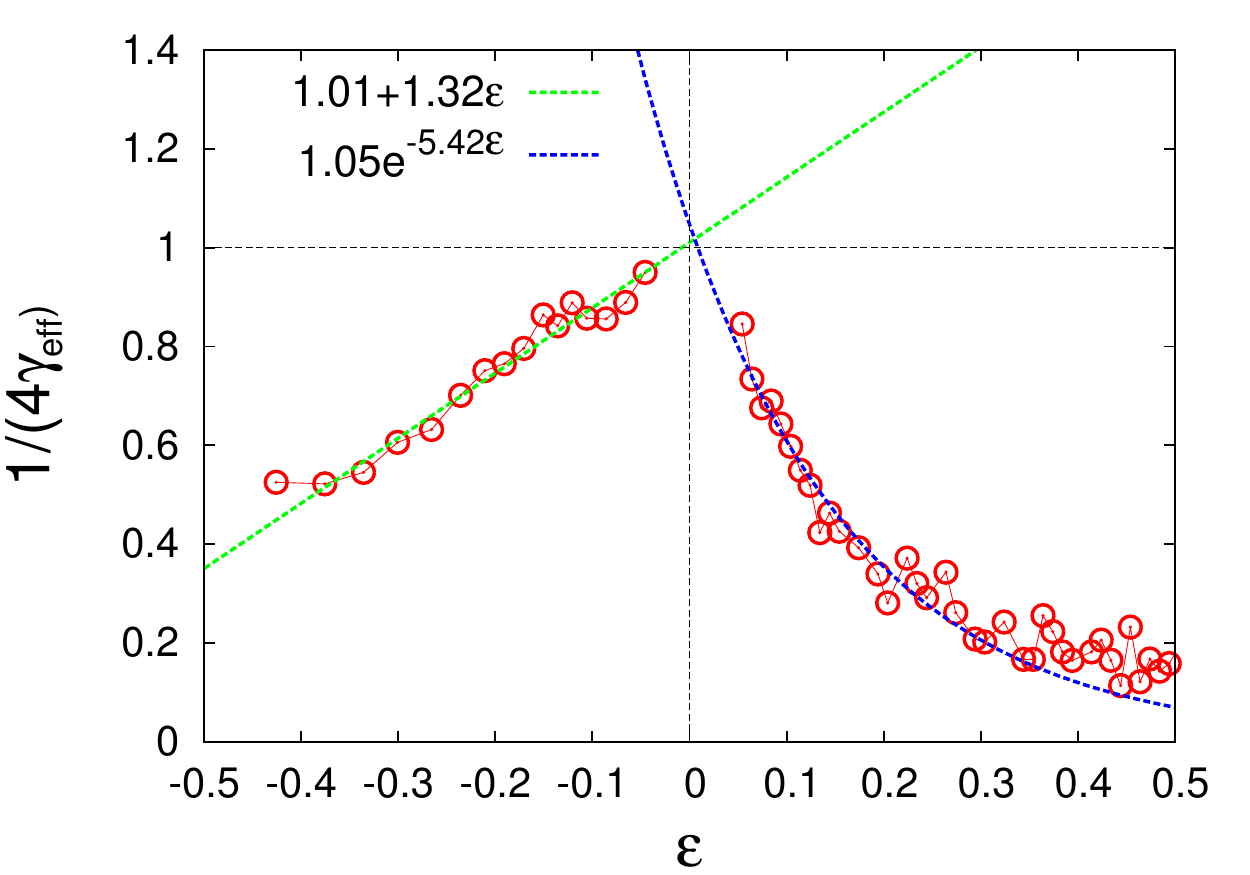}
\caption{(Color online)
$1/(4\gamma_{\rm{eff}})$ versus $\epsilon$ for $N=6400$.
Two dotted lines represent the fitting lines,
indicating $\gamma\approx 0.24$ and $\gamma' \approx 0.25$.}
\label{fig:gammaeff_regular}
\end{figure}

We also estimate $\gamma$ and $\gamma'$ directly from off-critical data by examining the effective exponents $\gamma_{\rm{eff}}$
for large $N$. Figure \ref{fig:gammaeff_regular} shows  $(4\gamma_{\rm{eff}})^{-1}$ versus $\epsilon$.
Following a similar procedure as in the random case, we find $\gamma=0.24(2)$ and $\gamma'=0.25(1)$.
We also fit the off-critical data directly in Fig.~\ref{fig:chi_regular}, which
agree very well with $\chi\sim |\epsilon|^{ -0.25}$.
All the numerical results strongly suggest scaling exponent values as
\begin{equation}
\beta=\frac{1}{2}, \quad \bar\nu=\bar\nu'=\frac{5}{4},  \quad \text{and} \quad \gamma=\gamma'=\frac{1}{4} ~.
\label{eq:conjecture_regular}
\end{equation}
Our results are clearly different from Daido's numerical suggestions of $\bar\nu\approx \frac{1}{2}$ and $\bar\nu'\approx 2$
and also his theoretical prediction of $\gamma^\prime=1$ on the subcritical side, which cannot be accounted for
by adopting a slightly different definition of $\chi$ by Daido\cite{ref:Daido-ex}.

\subsection{Hyperscaling relation}

The hyperscaling relation, $\gamma=\bar\nu-2\beta$, is seen to hold in the regular case.
Furthermore, there is no detectable ``disorder fluctuation'' arising from different initial phase values, leading to
$\left[\langle\Delta\rangle\right]^2\approx \left[\langle\Delta\rangle^2\right]$.

\section{Summary}

In this paper, we revisited the well-known Kuramoto model  and investigated the FSS of
the order parameter and its dynamic fluctuations near the onset of
synchronization transition. We find that critical scaling behavior crucially depends
on the presence of disorder among intrinsic frequencies of oscillators. In the random
case with disorder, it is shown analytically and numerically that the FSS exponent $\bar\nu=\frac{5}{2}$ and the dynamic
fluctuation exponent $\gamma\approx 1$, in both the supercritical and the subcritical region.
In the regular case without disorder, we find $\bar\nu\approx \frac{5}{4}$ and $\gamma\approx \frac{1}{4}$.
In both cases, the exponent $\bar\nu$ is different from its conventional value $\bar\nu=2$ in usual mean-field theory
of homogeneous systems with global coupling.

In the random case, disorder in the oscillator frequencies broadens the critical region and drives order parameter
fluctuations in a sample-dependent manner. The hyperscaling relation that describes the interplay between
static (i.e. averaged) and dynamic order parameter fluctuations is violated.
Interestingly, a modified mean-field theory that completely ignores dynamic fluctuations yields correct
critical FSS of the order parameter, although the amplitude is renormalized by dynamic fluctuations.
This is one of the main findings in this work. In comparison, numerically determined exponents $\gamma$ and
$\bar\nu$ in the regular case do satisfy the hyperscaling relation. Further work is required to explain the
particular values observed in our study.

\section*{Acknowledgments}
This research was supported by the NRF Grants No. 2012R1A1A2003678 (H.H) and 2013R1A1A2A10009722 (H.P),
and by the NSFC Grant No. 11175013 (L.-H.T).
We thank Korea Institute for Advanced Study for providing computing resources for this work.

\appendix
\section{Finite-size scaling in the mean-field approximation}

In this appendix we consider in some detail the solution to the self-consistent mean-field equation
(\ref{eq:SCE-FS}). For this purpose, it is convenient to introduce a function
\begin{equation}
\psi(u)=\left\{
\begin{array}{ll}
\sqrt{1-u^2}, &  |u|<1; \\
0, & |u|\geq 1.
\end{array}
\right.
\label{eq:psi-def}
\end{equation}
In terms of $\psi$, we have,
\begin{equation}
\Psi(K\Delta)={1\over N}\sum_{j=1}^N\psi(\omega_j/K\Delta).
\label{eq:Psi_new}
\end{equation}
Since each term in the sum is a monotonically increasing function of $z=K\Delta$, independent of the
sign of $\omega_j$, the sum is also a monotonically increasing function of $z$. In terms of $z$,
Eq. (\ref{eq:SCE-FS}) can be rewritten as,
\begin{equation}
K^{-1}z=\Psi(z).
\label{eq:self-consistent}
\end{equation}
When $K$ is sufficiently small, only the trivial solution $z=0$ is obtained. The nontrivial solution
is obtained only for $K>K_c$ where $K_c$ depends on the specific choice of the frequencies
$\{\omega_j\}$.

We now consider the statistical properties of the solution to Eq. (\ref{eq:self-consistent}) when the
frequencies $\{\omega_j\}$ are drawn independently from a given distribution $g(\omega)$.
In this case,  $\psi_j\equiv\psi(\omega_j/z)$ are also independent random variables whose
distribution can be obtained from $g(\omega)$. The quantity $\Psi$ as defined by Eq. (\ref{eq:Psi_new})
is simply the average of the random variables $\psi_j, j=1,\ldots, N$. Its mean value is given by $[\psi]$.
Let $\delta\Psi=\Psi-[\psi]$. Its variance is given by
\begin{equation}
[(\delta\Psi)^2]={[\psi^2]-[\psi]^2\over N}.
\label{eq:var-Psi}
\end{equation}
The variance of $\psi$ can be easily computed from $g(\omega)$:
\begin{eqnarray}
[\psi^2]-[\psi]^2
&=&\int_{-z}^z d\omega g(\omega)\Bigl(1-\bigl({\omega\over z}\bigr)^2\Bigr)\nonumber\\
&&\quad -\Bigl(\int_{-z}^z d\omega g(\omega)\sqrt{1-\bigl({\omega\over z}\bigr)^2}\ \Bigr)^2\nonumber\\
&=&{4\over 3}g(0)z - \Bigl({\pi\over 2}g(0)\Bigr)^2z^2+ ...
\label{eq:varpsi}
\end{eqnarray}
Invoking the central limit theorem, we conclude that
the distribution of  $\Psi$ tends to a Gaussian at large $N$, with mean
and variance as stated above.

For small $z$,
\begin{eqnarray}
[\psi]&=&\int_{-z}^z d\omega g(\omega)\sqrt{1-\bigl({\omega\over z}\bigr)^2}\nonumber\\
&=&{\pi\over 2}g(0)z+{\pi\over 16}g''(0)z^3+\ldots
\label{eq:psi-mean}
\end{eqnarray}
Equation (\ref{eq:self-consistent}) can now be written in the form
\begin{equation}
K^{-1}z=K_c^{-1}z -cz^3 +d\Bigl({z\over N}\Bigr)^{\frac{1}{2}}\eta,
\label{eq:mf-fluc}
\end{equation}
where $K_c^{-1}=\pi g(0)/2$, $c=-\pi g''(0)/16>0$, and $d=(4g(0)/3)^{\frac{1}{2}}$.
Here $\eta=\delta\Psi/[\delta\Psi^2]^{\frac{1}{2}}$ is a Gaussian random variable with zero mean and unit variance.

Solution to Eq. (\ref{eq:mf-fluc}) can be cast
in the scaling form (\ref{eq:Delta_random_FSS}) with $\beta=\frac{1}{2}$ and $\bar\nu=\frac{5}{2}$.
Specifically, let $z=K_cN^{-\frac{1}{5}}f$, Eq. (\ref{eq:mf-fluc}) becomes
\begin{equation}
xf-cK_c^3f^3+\Bigl({8\over 3\pi}\Bigr)^{\frac{1}{2}}f^{\frac{1}{2}}\eta =0,
\label{eq:f-eta}
\end{equation}
where $x=N^{\frac{2}{5}}(K-K_c)/K$ is the scaled distance to the transition point.

We now consider solution to Eq. (\ref{eq:f-eta}) at $K=K_c$ or $x=0$. Simple algebra yields
\begin{equation}
f(0)=\left\{
\begin{array}{ll}
A\eta^{\frac{2}{5}}, &  \eta>0; \\
0, & \eta\leq 0.
\end{array}
\right.
\label{eq:f-eta2}
\end{equation}
Here $A=(8/3\pi)^{\frac{1}{5}}c^{-\frac{2}{5}}K_c^{-\frac{6}{5}}$. The $n$th moment of $f(0)$ at the transition is given by ($n>0$),
\begin{eqnarray}
[f^n(0)]&=&{1\over (2\pi)^{\frac{1}{2}}}\int_0^\infty d\eta e^{-\eta^2/2}A^n\eta^{2n/5}\nonumber\\
&=&{A^n\over (2\pi)^{\frac{1}{2}}}\int_0^\infty dt e^{-t}(2t)^{n/5-\frac{1}{2}}\nonumber\\
&=&{2^{n/5}\over 2\pi^{\frac{1}{2}}}A^n\Gamma\Bigl({n\over 5}+{1\over 2}\Bigr).
\label{eq:f-n}
\end{eqnarray}

In the special case $g(\omega)=e^{-\omega^2/2}/\sqrt{2\pi}$, $A=(8\pi/3)^{\frac{1}{5}}$. The first four moments
are then given by
\begin{eqnarray}
[f(0)] &=&{1\over 2\pi^{\frac{1}{2}}}\Bigl({16\pi\over 3}\Bigr)^{\frac{1}{5}}\Gamma\bigl({7\over 10}\bigr)=0.643454..,\nonumber\\
\ [f^2(0)] &=&{1\over 2\pi^{\frac{1}{2}}}\Bigl({16\pi\over 3}\Bigr)^{\frac{2}{5}}\Gamma\bigl({9\over 10}\bigr)=0.930853..,\nonumber\\
\ [f^3(0)]&=&{1\over 2\pi^{\frac{1}{2}}}\Bigl({16\pi\over 3}\Bigr)^{\frac{3}{5}}\Gamma\bigl({11\over 10}\bigr)=1.45621..,\nonumber\\
\ [f^4(0)]&=&{1\over 2\pi^{\frac{1}{2}}}\Bigl({16\pi\over 3}\Bigr)^{\frac{4}{5}}\Gamma\bigl({13\over 10}\bigr)=2.41398...\nonumber
\end{eqnarray}


\begin{thebibliography}{20}
\bibitem{ref:Daido-ex} In references~\cite{ref:Daido}, a slightly different definition for dynamic fluctuations
is adopted as $N\left[\langle |Z|^2\rangle-|\langle Z\rangle|^2\right]$ with the complex order parameter $Z$
in Eq.~(\ref{eq:def_Delta}). Given that the dynamic equations (\ref{eq:model}) have the symmetry of
$\phi_j\rightarrow\phi_j+{\rm constant}$ for all $j$, this quantity behaves in the same way as $N\left[\langle\Delta^2\rangle\right]$
in the absence of symmetry breaking, which is expected of a finite system.
\end{thebibliography}
\end{document}